\documentclass[onecolumn,aps,pre]{revtex4-1}
\usepackage{amsmath}
\usepackage[utf8]{inputenc}
\usepackage{setspace}
\usepackage{caption}
\usepackage{color}
\usepackage{subcaption}
\usepackage{graphicx}
\usepackage{color}

\begin{document}
\title{Front Propagation and Clustering in the Stochastic Nonlocal Fisher Equation}
\author{Yehuda A. Ganan, David A. Kessler}
\affiliation{Dept. of Physics, Bar-Ilan University, Ramat-Gan 52900 Israel}
\begin{abstract}
The nonlocal Fisher equation is a diffusion-reaction equation where the reaction has a  linear birth term and a nonlocal quadratic competition, which describes the reaction between distant individuals.
This equation arises in evolutionary biological systems, where the arena for the dynamics is trait space, diffusion accounts for mutations and individuals with similar traits compete, resulting in partial niche overlap. It has been found that the (non-cutoff) deterministic system gives rise to a spatially inhomogeneous state for a certain class of interaction kernels, while the stochastic system produces an inhomogeneous state for small enough population densities. Here we study the problem of front propagation in this system, comparing the stochastic 
dynamics to the heuristic approximation of this system by a deterministic system where the linear growth term is cut off below some critical density.  Of particular interest is the nontrivial pattern left behind the front.  
For large population density, or small cutoff, there is a constant velocity wave propagating from the populated region to the unpopulated region. As in the local Fisher equation, the spreading velocity is much lower than the Fisher velocity which is the spreading velocity for infinite population size. The stochastic simulations give approximately the same spreading velocity as the deterministic simulation with appropriate birth cutoff. 
When the population density is  small enough, there is a different mechanism of population spreading. The population is concentrated on clusters
which divide and separate. This mode of spreading has small spreading
velocity, decaying exponentially with the range of the interaction kernel. The dependence on the carrying capacity is more complicated, and the log of the velocity scales as a power law of the carrying capacity, where the power is dependent on the kernel.
We also discuss the transition between the bulk homogeneous pattern to separated islands, which occurs when the
minimal population density is lower than the cutoff in the deterministic model.  
\end{abstract}
\maketitle

\section{Introduction}
Much study has been devoted to the effects of demographic stochasticity in spatial birth-death processes.  These effects are very apparent in evolutionary models, where the speed of evolution is controlled by the finite population size.  The primary mechanism underlying this is a clustering effect, where the individuals are strongly localized in trait space, thereby limiting the variance and hence the speed of evolution.  This clustering effect is also seen in an even simpler context, that of a fixed number of migrating individuals undergoing birth and death, where all the individuals are localized to a region of size proportional to the square root of the population size.  Recently, McKane and coworkers~\cite{rossberg2013there} showed that a clustering effect is also present in a birth-death model where the competition term limiting the population size is of finite range.  They suggested that such an effect might be an important mechanism driving sympatric speciation.  The goal of this current work is to further explore this phenomenon, and show how their model relates to other classic models, such as the models mentioned above as well as the famous Fisher model of front propagation~\cite{fisher1937wave,kolmogorov1937etude}.

These problems are difficult to analyze due to their nonlinear nature.
At the simplest level, there are three types of processes occurring in the models: birth, death and mutation.
The birthrate is linear in the current population. Mutations induce a difference between the parent and its descendant.
If the mutation rate is relatively small, which is usually the case, it can be described by a diffusion process. Therefore
these two processes are linear processes. The death process contains a linear part, but it also contains a nonlinear part
that emerges from the competition for resources between individuals. Besides being nonlinear, the competition process is typically nonlocal
 in trait space. In the simplest model of an evolving population, for example, the competition is global, with the overall population kept fixed, so that all individuals fight against all others for a slot in the population. More generally, however, the competition is of finite range, meaning that individuals sufficiently distant experience weakened competition, due to reduced niche overlap.
 
The extreme global limit of infinite-range competition has been considered by many different authors~\cite{zhang1990diffusion,meyer1996clustering,kesslerjsp,young2001reproductive,houchmandzadeh2002clustering,rogers2012demographic}, with the most catchy name for the model being ``Brownian bugs". Here we have a fixed finite population, of size $N$, undergoing birth (at rate $r$), death and mutation (or equivalently diffusion).  As mentioned above, this  system exhibits a striking phenomenon, namely clustering.  The entire population stays together, with the average distance between all pairs of particles converging to a time-independent value of order $\sqrt{rD N}$, where $D$ the diffusion (mutation) rate. This is clearly a finite population size effect, since diffusing entities described by the continuum diffusion equation are characterized by a distance scale that increases unboundedly with $t$ as $\sqrt{D t}$.  The simplest evolution models are just an extension of this model wherein the birthrate is a function of position, and so the cluster moves to regions of higher birthrate~\cite{tsimring1996RNA}.
 
The other extreme limit, namely of purely local competition, has also been extensively studied.  Here, the stable steady-state is trivial, with constant average population density.  The dynamics is more interesting, in that an initial bounded populated region spreads in time, with average constant velocity, independent of the details of the initial condition. 
In the infinite population limit, these systems can be described by a differential equation for the development in time of the initial population density. 
The first ones to discuss such systems were Fisher~\cite{fisher1937wave} and Kolmogorov, Petrovsky and Piskunov~\cite{kolmogorov1937etude}
back in 1937.  There is  a certain velocity $v_{\text{Fisher}}$ such that there is a non-negative propagating wave 
solution for any velocity $v\ge v_{\text{Fisher}}$. 
If  initially  the population is confined to some compact region the solution that emerges is this of a front propagating
with the minimal velocity, $v_{\text{Fisher}}$. 
The selection mechanism of the minimal velocity solution from all the other possible solution was discovered by Aronson and Weinberger 
in the mid 70s~\cite{aronson1975nonlinear}.
Stochastic simulations for large but finite population show that the average velocity is lower than $v_{\text{Fisher}}$ and converges very slowly to 
$v_{\text{Fisher}}$ as the population is increased. The main reason for the difference was explained by Brunet and Derrida~\cite{brunet1997shift}. They showed that the difference
emerges from the effective introduction of a birth cutoff to the tail of the front, where the population density in the continuous model is exponentially small. In this cutoff model, the
achieved velocity converges to the continuous velocity like $\ln(N_C)^{-2}$, where no births are allowed when the local population density falls below $1/N_C$.

Our focus in this paper is on connecting these two well-understood limits, i.e. considering the case with finite interaction range, $w$. This problem has also received much attention~\cite{rossberg2013there,sasaki1997clumped,fuentes2003nonlocal,hernandez2004clustering,maruvka2006nonlocal}.  In the infinite population limit, the dynamics is described by a nonlocal equation of the form:
\begin{equation}
\label{eq:gov}
\dfrac{d\rho(x)}{dt} = a\rho(x)+D\dfrac{d^2\rho(x)}{dx^2}-\dfrac{a\,\rho(x)}{N_C}
\int K\left(\frac{x-x'}{w}\right)\rho(x')dx'
\end{equation}
where $\rho$ is the local population density, $a$ is a the time scale of the birth process, $D$ is the diffusion constant and $N_C$ is a parameter which sets the scale of the population in the range $w$.
This equation has a homogeneous populated state where $\rho(x)=\rho_c$ and this solution is stable if and only if $\forall\,k:-D\,k^2-a\,w^2{\widetilde{K}(k)}<0$ 
where $\widetilde{K}$ is the Fourier transform of the kernel $K$.
Since $k^2$ is always positive, we get that if $\widetilde{K}(k)$ is always positive then the homogeneous solution is stable. 
In particular,  the homogeneous solution is stable for a Gaussian kernel.
Equation \ref{eq:gov} may have other non-homogeneous solutions, which cannot be found analytically. However, Berestycki et al.~\cite{berestycki2009non} proved that this kind
of solutions does not exist if $\widetilde{K}(k)$ is always positive, or if $w$ is small enough.  
  As with the case of the $w\to \infty$, Brownian bugs, limit, the
steady-state of the stochastic process can be markedly different from that of the infinite population limit.
McKane, Rossberg and Rogers~\cite{rossberg2013there} found out that for low enough $N_C$,  a spatially inhomogeneous state may arise, even in cases where the homogeneous solution is stable. This pattern takes the  form of separated concentrations of populations.
 
 As mentioned above, the local Fisher-KPP equation is the limit of $w\to0$. 
Traveling wave solutions also exist in the nonlocal equation, and Berestycki et al.~\cite{berestycki2009non} proved that this equation  has a set of traveling wave solutions for each $v\ge v_{\text{Fisher}}=2\sqrt{aD}$. The spreading velocity for an initially bounded initial condition is just the Fisher velocity, since the nonlinearity is irrelevant in the velocity selection.   In this case, the propagating front leaves behind the inhomogeneous state, similar to the situation in the Swift-Hohenberg equation, with the wavelength of the pattern being fixed by the velocity selection.
The other limiting case is where $w\to\infty$.  Here, at the level of the deterministic equation, the competition term is just $\rho(x)$ times the overall population size divided by $N_c$. The population quickly saturates at $N_c$ at which point the competition term exactly cancels the growth term, and we are left with pure diffusion.  Thus, at the deterministic level, the spreading velocity is zero.  Here, as we have already indicated, the stochastic effects are crucial, and despite the spreading effect of diffusion, the particles remain clustered~\cite{meyer1996clustering,rogers2012demographic}.
The population clusters and moves together (diffuses) as a unit rather than spreading (even diffusively, as the deterministic equation would indicate) through space. The width of the cluster is $\sqrt{\dfrac{2D(N_c-1)}{a}}$.

This article will further examine the spatial propagation of the stochastic nonlocal Fisher equation, and how it relates to its two limits, the Fisher equation and the Brownian bug model. An important question is how this stochastic clustering, which at its heart is independent of the nature of the interaction kernel $K$, is impacted by the deterministic clustering present for some interaction kernels. We will focus on two kernels: Gaussian kernel $K(x)=\frac{1}{\sqrt{2\pi}}\exp(-\frac{x^2}{2})$ which is
stable in the continuous limit, and flat-top kernel $K(x)=0.5\,\theta( 1- \lvert x \rvert )$ which does exhibit deterministic clustering. The scheme of this paper is as follows. In the next section we will describe the simulation methods and present some simulations.  We will then study the spreading velocity of the population for two kernels in the limits of high population and of low population. In the last section we will examine the details of the process by which separated species are created.

\section{Simulation Methods}
The deterministic equation is simulated using finite difference methods and Euler integration. 
For small population (i.e. small $N_C$) stochastic simulations can be done by simulating each individual. Each individual diffuses independently and
in each time step we compute the death rate of the $i$-th individual $r_D =N_C^{-1} \cdot \sum_{j=1}^{N(t)}K\left(\frac{x_i-x_j}{w}\right)$,  where $N(t)$
is the number of individuals at time $t$.
The chance of each individual to die in the time step is $r_D\,dt$. Each individual gives birth independently  
at a rate $a$, and so the number of births each individual has in time interval $dt$ is a Poisson random variable with mean $a\,dt$.

For large population sizes, the method described above is inefficient. In order to simulate the stochastic system more effectively, we bin the space.  
The status of the system is described by the set $\left\{N_i\right\}_{i=1}^{N_x}$ where $N_x$ is the number of spatial bins and $N_i$ is the total population in each bin. 
The total number of births in each bin in time interval $dt$ is a Poisson random variable with mean $N_i\,a\,dt$ 
The death rate of each individual inside the $i$-th bin is $r_D=N_C^{-1} \cdot \left(\sum_{j=1}^{N_x}K\left(\frac{x_i-x_j}{w}\right)N_j \right)$, 
The number of deaths in each bin at a time step is described by a random number taken from Binomial distribution with parameters $\left(N_i,r_D\,dt\right)$.
The difference between the death and birth processes is caused by the fact that each individual can give multiple births at any time interval, but can die only once. 
The diffusion process is described by binomial random numbers, where we only allow diffusion to a neighbor cell, 
and the probability of each individual to move to a neighbor cell is $\widetilde{D} = \dfrac{Ddt}{dx^2}$. 
The number of individuals moving to the $i-1$-th cell from the $i$-th cell ($L_i$) is taken from a binomial distribution with parameters 
$\left(N_i,\widetilde{D}\right)$ and the number of individuals moving to the $i+1$-th cell is taken from a binomial distribution with parameters 
$\left(N_i-L_i,\dfrac{\widetilde{D}}{1-\widetilde{D}}\right)$.
This gives the correct probability for the diffusion process. 
From the set ${N_i}$ we can calculate the population density by $\rho=\frac{N}{dx}$.

\begin{figure*}[!h]
	\centering
	\begin{subfigure}{0.48\textwidth}
		\includegraphics[width=\linewidth]{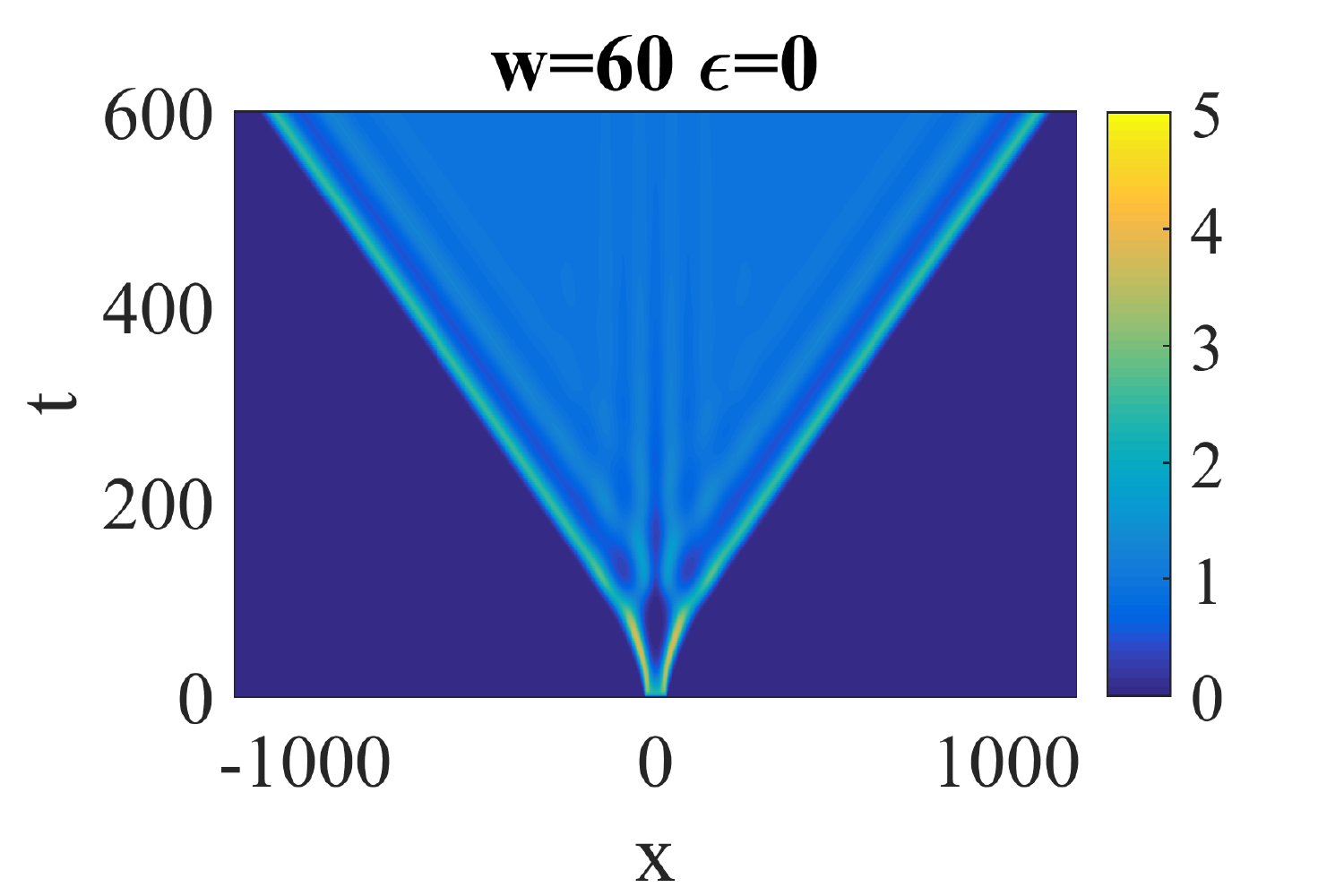}
	\end{subfigure}\hspace*{\fill}
	\begin{subfigure}{0.48\textwidth}
		\includegraphics[width=\linewidth]{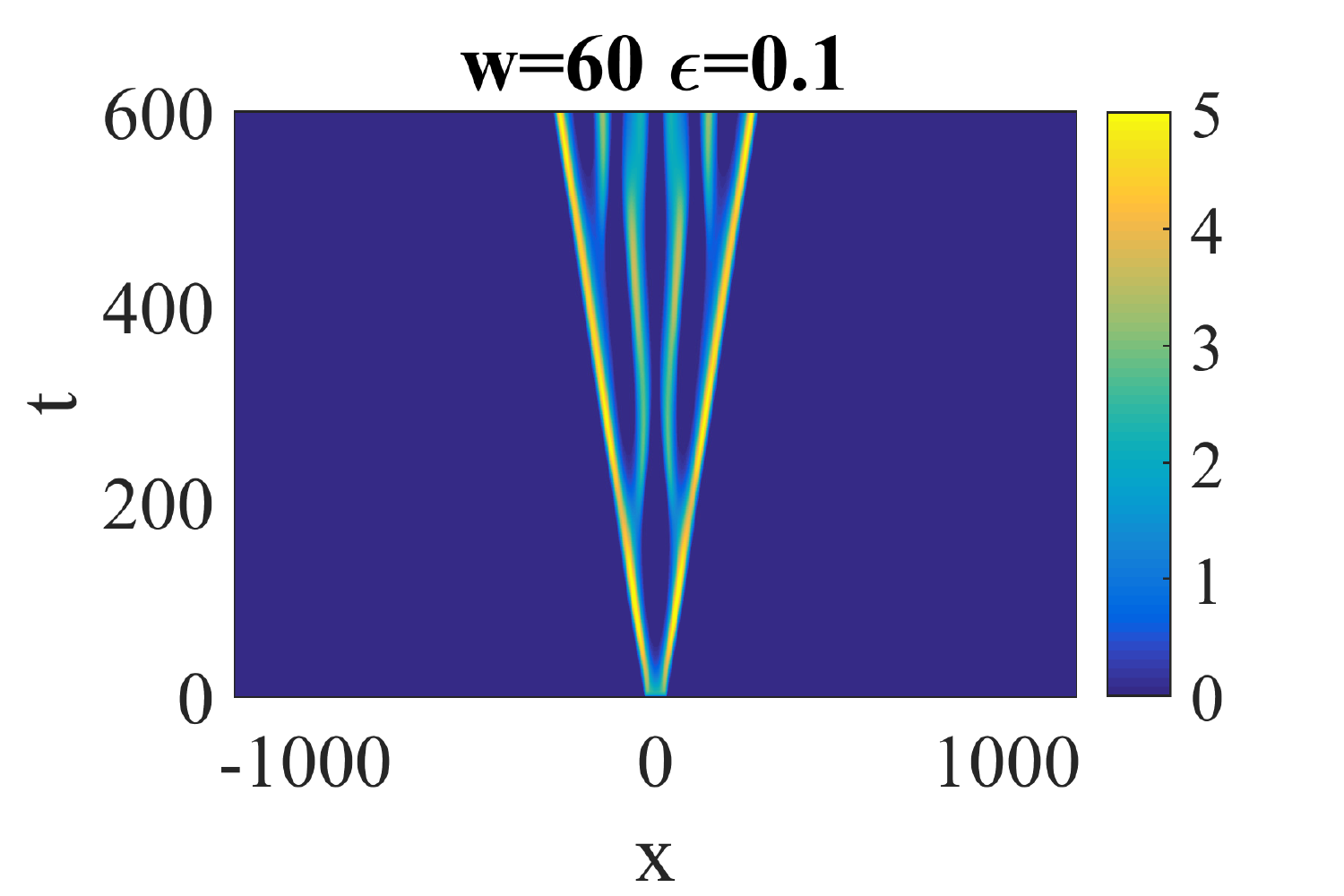}
	\end{subfigure}
	\medskip
	\begin{subfigure}{0.48\textwidth}
		\includegraphics[width=\linewidth]{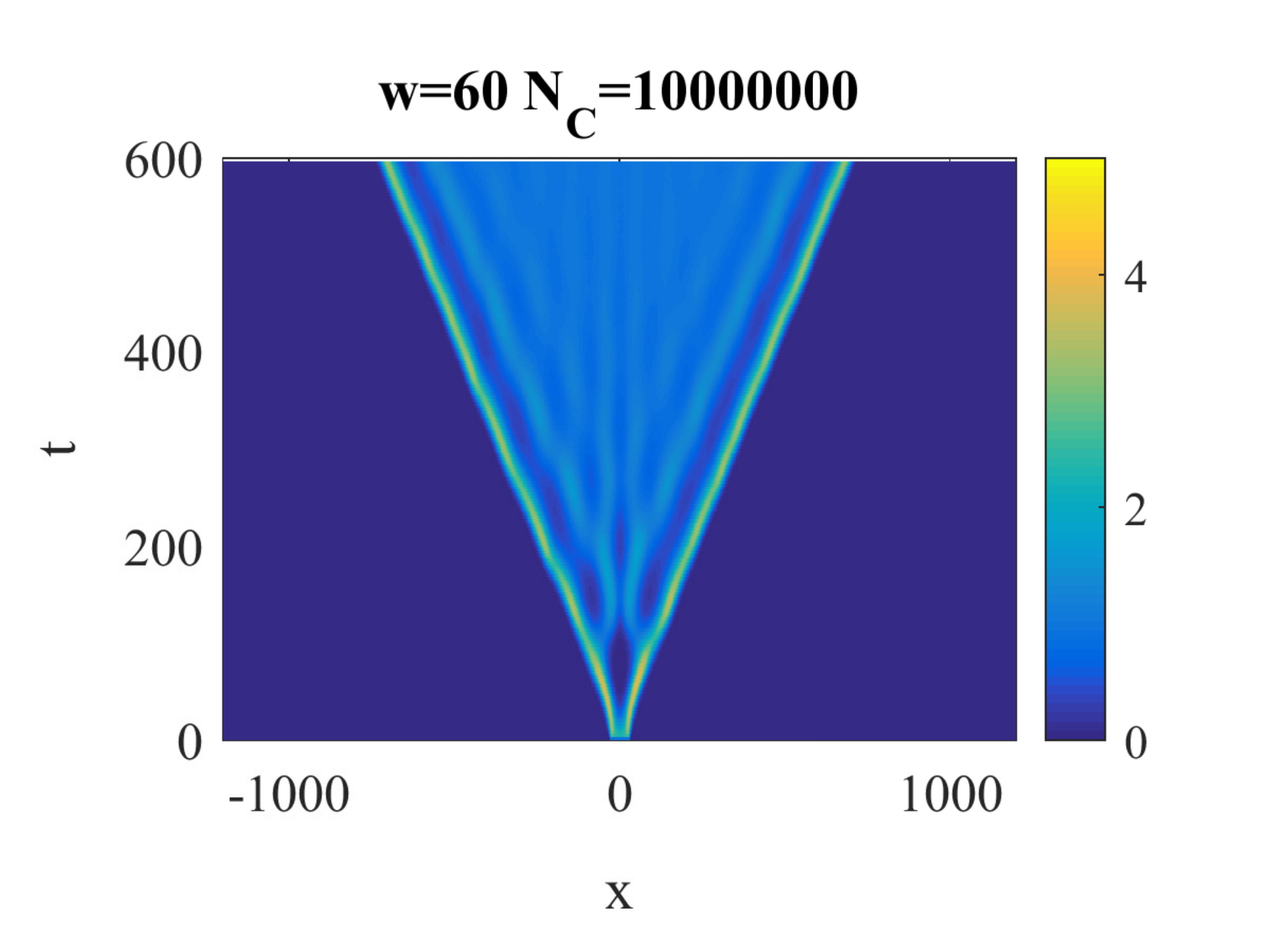}
	\end{subfigure}\hspace*{\fill}
	\begin{subfigure}{0.48\textwidth}
		\includegraphics[width=\linewidth]{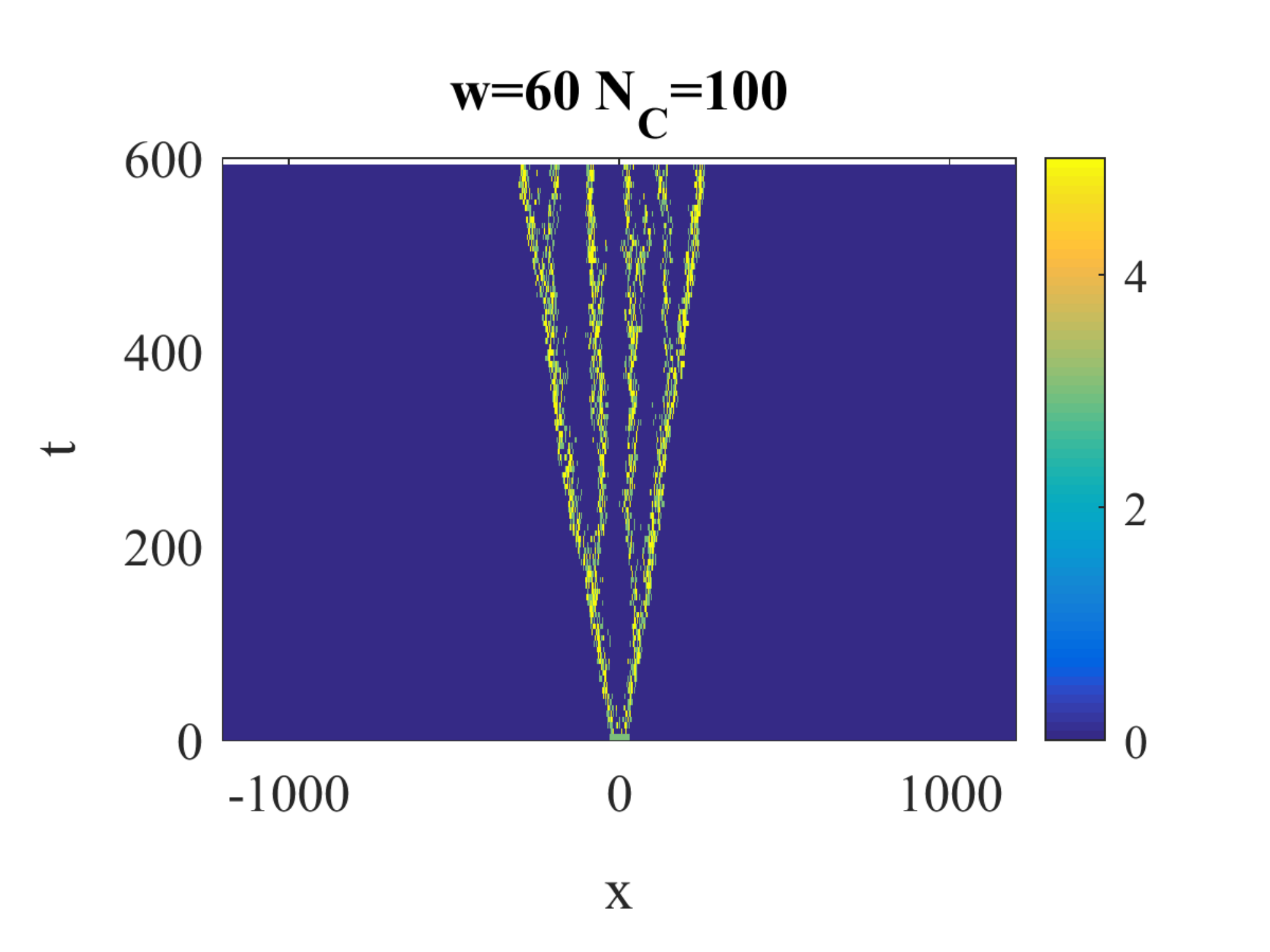}
	\end{subfigure}
	\caption{The normalized population density $\frac{\rho}{\rho_c}$ vs. time and space for $w=60$ and different values of $ \epsilon$ for the Gaussian kernel.  $D=a=1$}
	\label{fig:devGdet60}
\end{figure*}

\section{Spreading Velocity}
Consider an initial condition where the population is bounded in space. 
The diffusion process tends to cause the population to spread through space. When the population starts to occupy some space it reproduces until the 
death process balances the birth process, and the population density is constant up to stochastic fluctuations. This population 
will continue to spread further. In the local Fisher-KPP equation this behavior results in a traveling wave solution with constant velocity 
$v_{\text{Fisher}}=2\sqrt{aD}$.
In the Full competition case, after the logistic growth stage, the population only diffuses, so the spreading velocity is zero. 
In the continuous limit for values of $w$ between $0$ to $\infty$ the spreading velocity is still $v_{\text{Fisher}}$. This arises from the 
fact that in the tail of the front, the nonlocal interaction is  second order in $\rho$ and therefore the Fisher-KPP case holds.
Berestycki et al. proved that no velocity below $v_{\text{Fisher}}$ is allowed~\cite{berestycki2009non}, and any 
waves propagating faster than it decays.

Just as with the local stochastic Fisher-KPP equation, one can approximate the effect of stochasticity on the velocity by imposing a cutoff on the birth process when the number of individuals on a site is less than same number of order unity.  We expect the same to hold for the stochastic nonlocal model.

\begin{figure}[!h]
	\centering
	\begin{subfigure}{0.48\textwidth}
		\includegraphics[width=\linewidth]{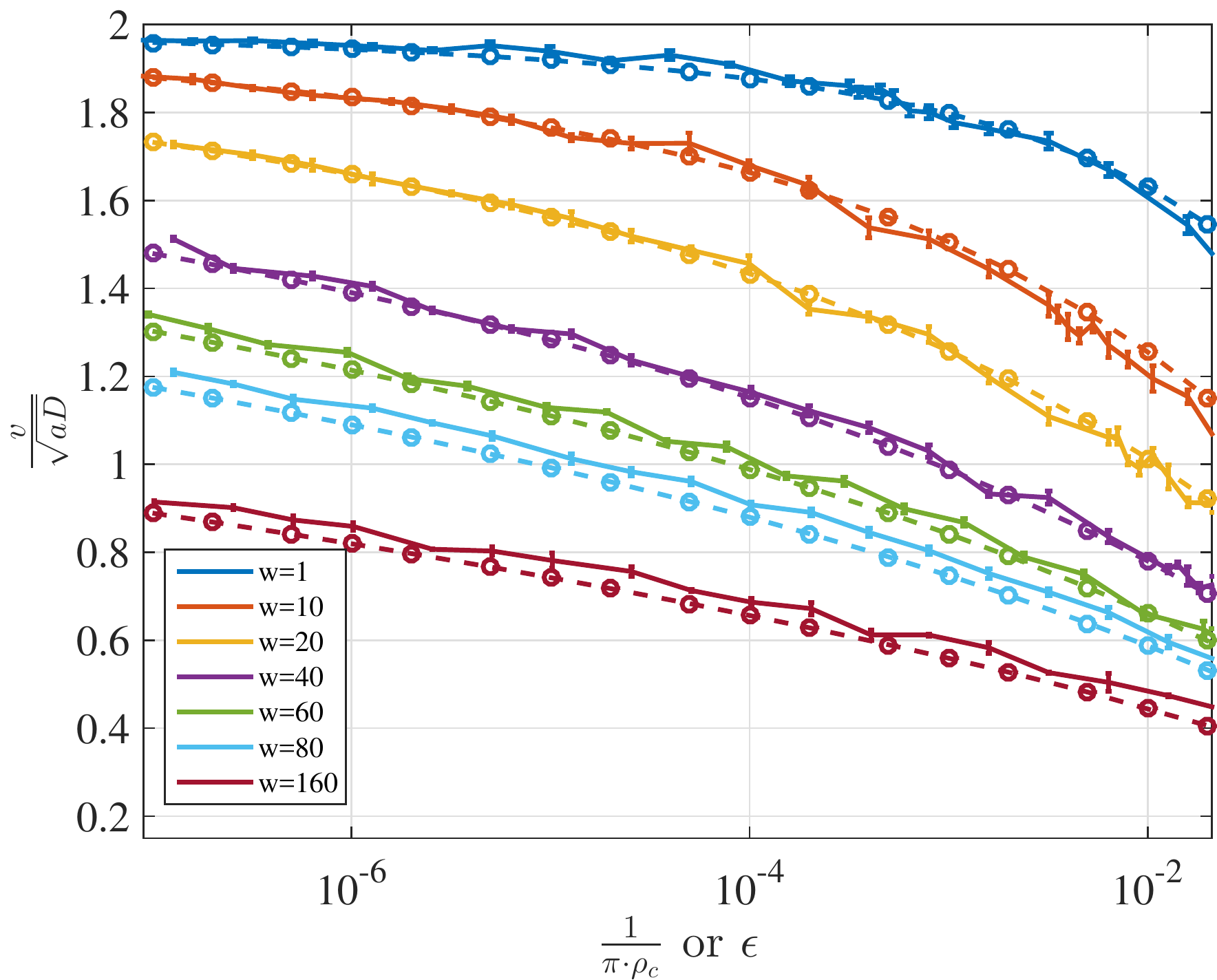}
	\end{subfigure}\hspace*{\fill}\\
	\begin{subfigure}{0.48\textwidth}
		\includegraphics[width=\linewidth]{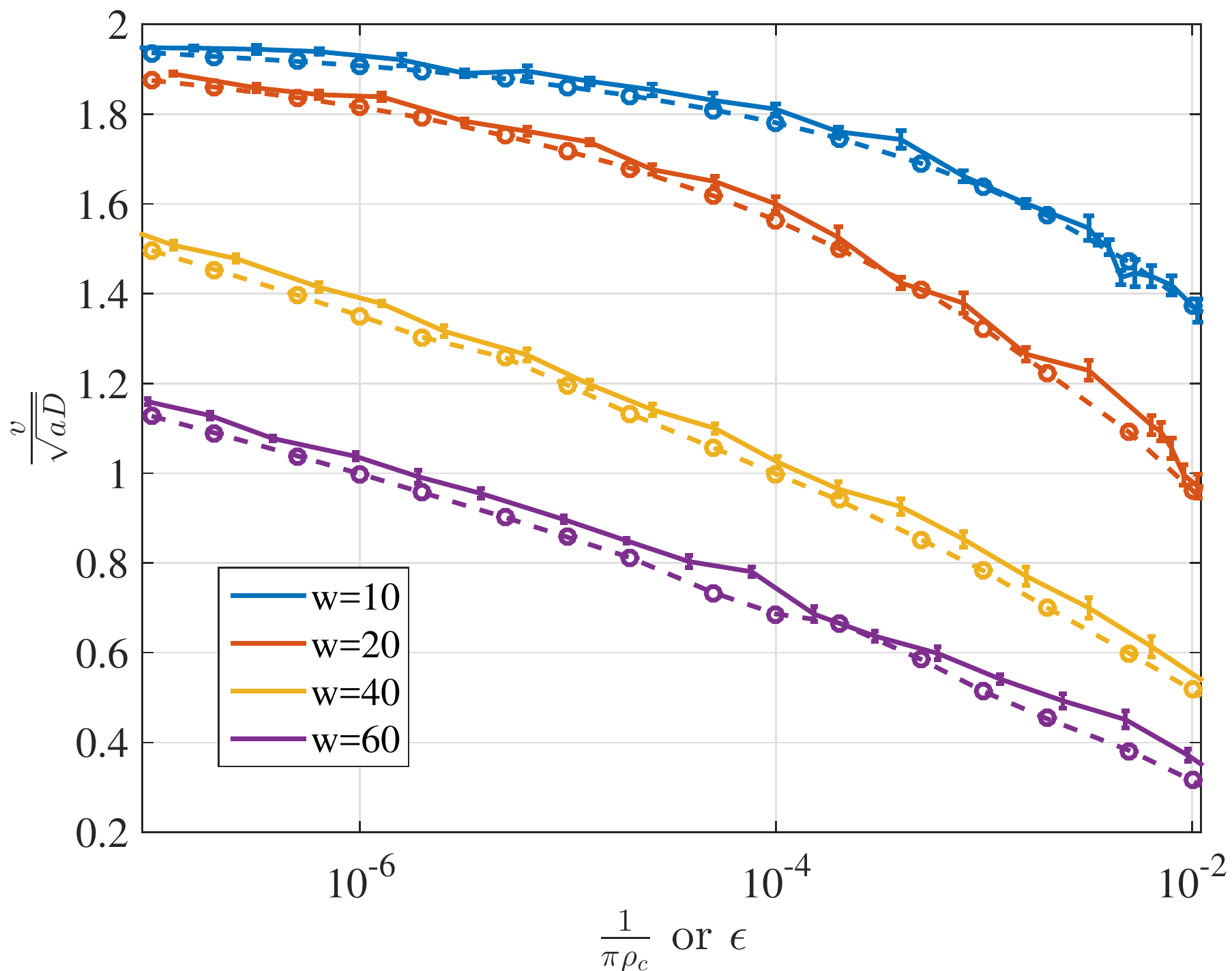}
	\end{subfigure}
	\caption{The spreading velocity vs. $N_C=\rho_c w$ and $\epsilon$ for different values of $w$. 
		Top - Gaussian kernel. Bottom - Flat-top kernel.
		Full lines - stochastic case. Dashed lines - deterministic case. $D=a=1$}
	\label{fig:velNeps}
\end{figure}

In Figure~\ref{fig:velNeps} we show the velocity against the birth cutoff $\epsilon$ for the (cutoff) deterministic case and versus $\frac{1}{N_C}$ for the stochastic case.
We can see good agreement between $v(\epsilon)$ to $v\left(\frac{1}{\pi\,\rho_c}\right)$, 
where $\pi$ serves as the phenomenological number of order unity connecting the cutoff to the local density. Interestingly, the agreement is good for both the Gaussian kernel and the flat-top kernel with this same factor of $\pi$.
\newline

\begin{figure}[!h]
	\centering
	\begin{subfigure}{0.48\textwidth}
		\includegraphics[width=\linewidth]{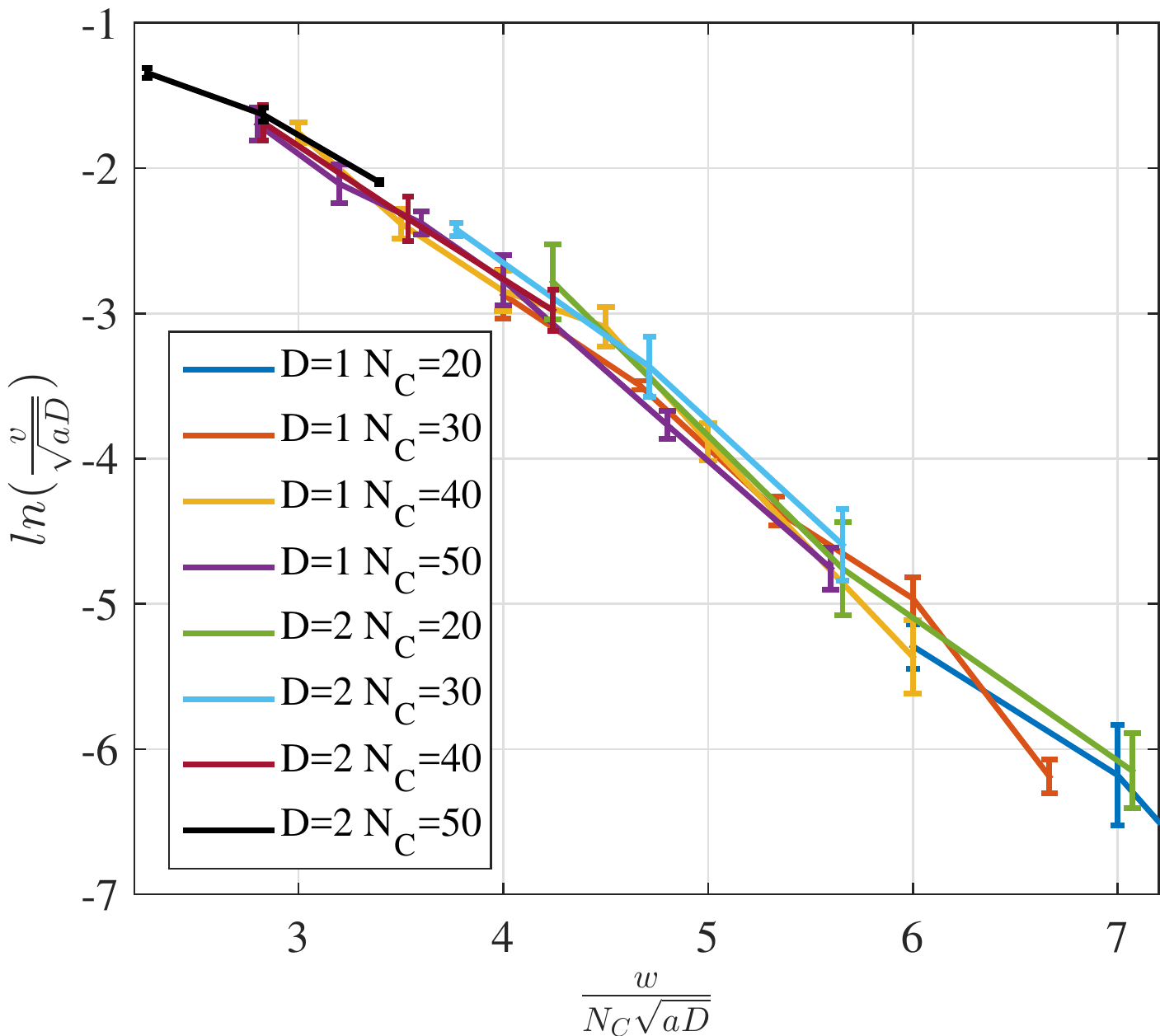}
	\end{subfigure}\hspace*{\fill}\\
	\begin{subfigure}{0.48\textwidth}
		\includegraphics[width=\linewidth]{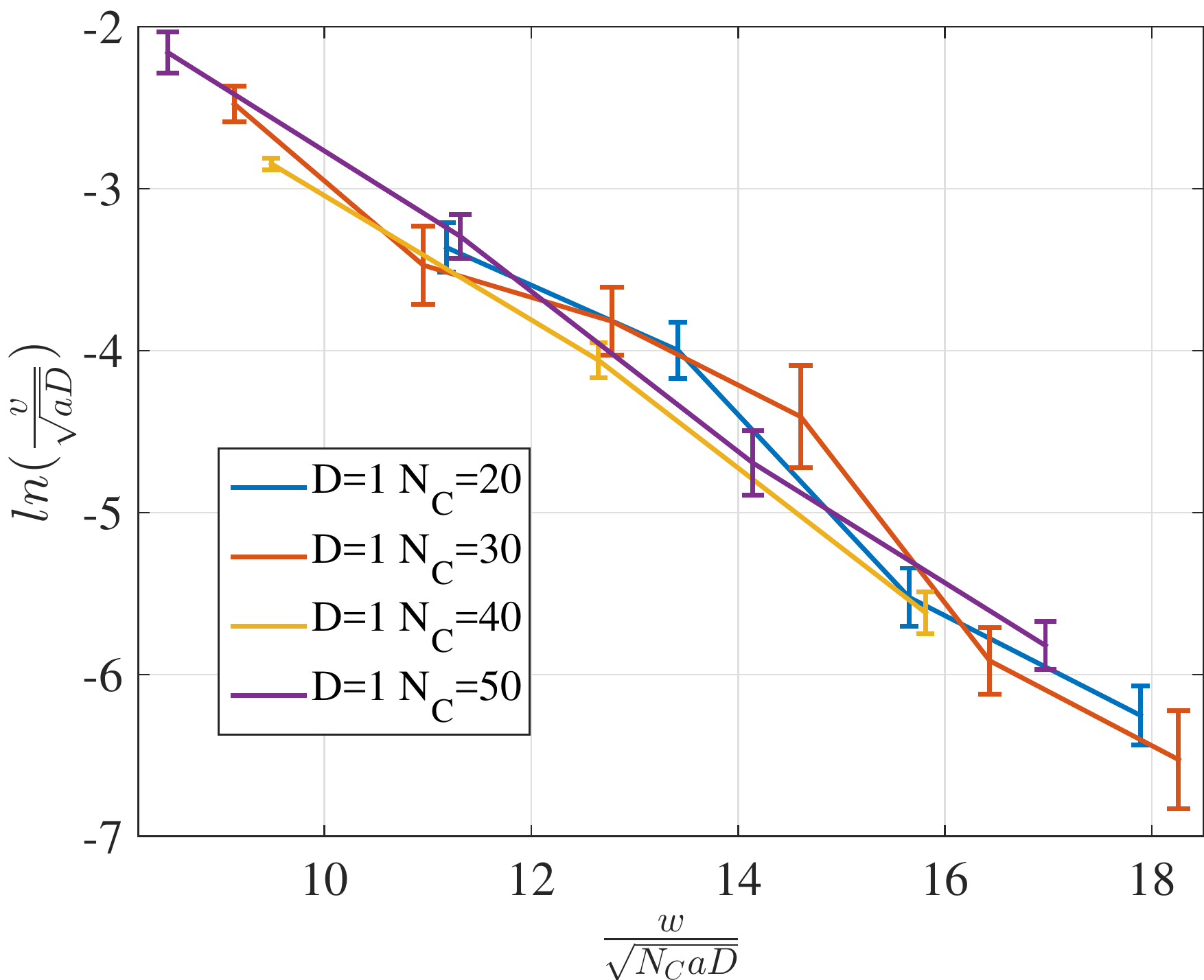}
	\end{subfigure}	
	\caption{The dependence of the spreading velocity on $N_C,w$ and $D$ for small population and large $w$ in the stochastic model.
		Top - Gaussian kernel. Bottom - Flat-top kernel. $a=1$}
	\label{fig:velsmall}
\end{figure}

For infinite $w$, there is a single cluster of fixed average size. There are fluctuations, however, exponentially rare, where the cluster has a much larger width.
Typically, this happens when the cluster temporarily splits in two.  The numbers in each subcluster fluctuate, and eventually, one of the clusters dies. However, for large but finite $w$, if the two subclusters manage to separate to a distance $w$, they no longer compete with each other and both grow to a full population size of $N_C$.  Since this depends on an exponentially rare fluctuation, we expect the spreading velocity in this case to be exponentially small. The expedient of introducing a cutoff into the equations cannot capture this effect, since it is truly a fluctuation driven phenomenon.  Rather, the cutoff equation predicts a single cluster of finite fixed size, as with Brownian bugs.  Figure~\ref{fig:velsmall} shows the dependence of the logarithm of the velocity on the parameters $N_C,w$ and $D$. 
The dependence on the parameter $N_C$ is apparently not the same for different kernels. for the flat-top kernel we get that $v\propto\exp(-N_C^{-0.5})$ and for Gaussian kernel 
$v\propto\exp(-N_C^{-1})$, other kernels shows the same features where $v\propto\exp(-N_C^{-\alpha})$, where $\alpha\approx1.2$ for a triangular kernel, 
$$K(x) = \theta\!\left(1-\lvert x \rvert\right)\left(1-\lvert x \rvert\right)$$ and 
$\alpha\approx0.7$ for a super-Gaussian kernel,
$$K(x) = \frac{\sqrt{2}}{\Gamma(0.25)}\exp(-x^4/4).$$
Overall we get that $v\propto\sqrt{aD}\exp\left(\dfrac{-bw\sqrt{a}}{N_C^{\alpha}\sqrt{D}}\right)$, which can be justified from dimensional analysis.

\section{Formation of Clusters}
In this section, we will investigate the processes leading to the formation of species in both the deterministic and the stochastic case. 
We will show that most of the processes are common to both cases, We will also discuss some attributes that comes from the noise in the stochastic
case and does not appear in the deterministic one.

\subsection{Gaussian Kernel}
\begin{figure*}[!ht]
	\centering
	\begin{subfigure}{0.48\textwidth}
		\includegraphics[width=\linewidth]{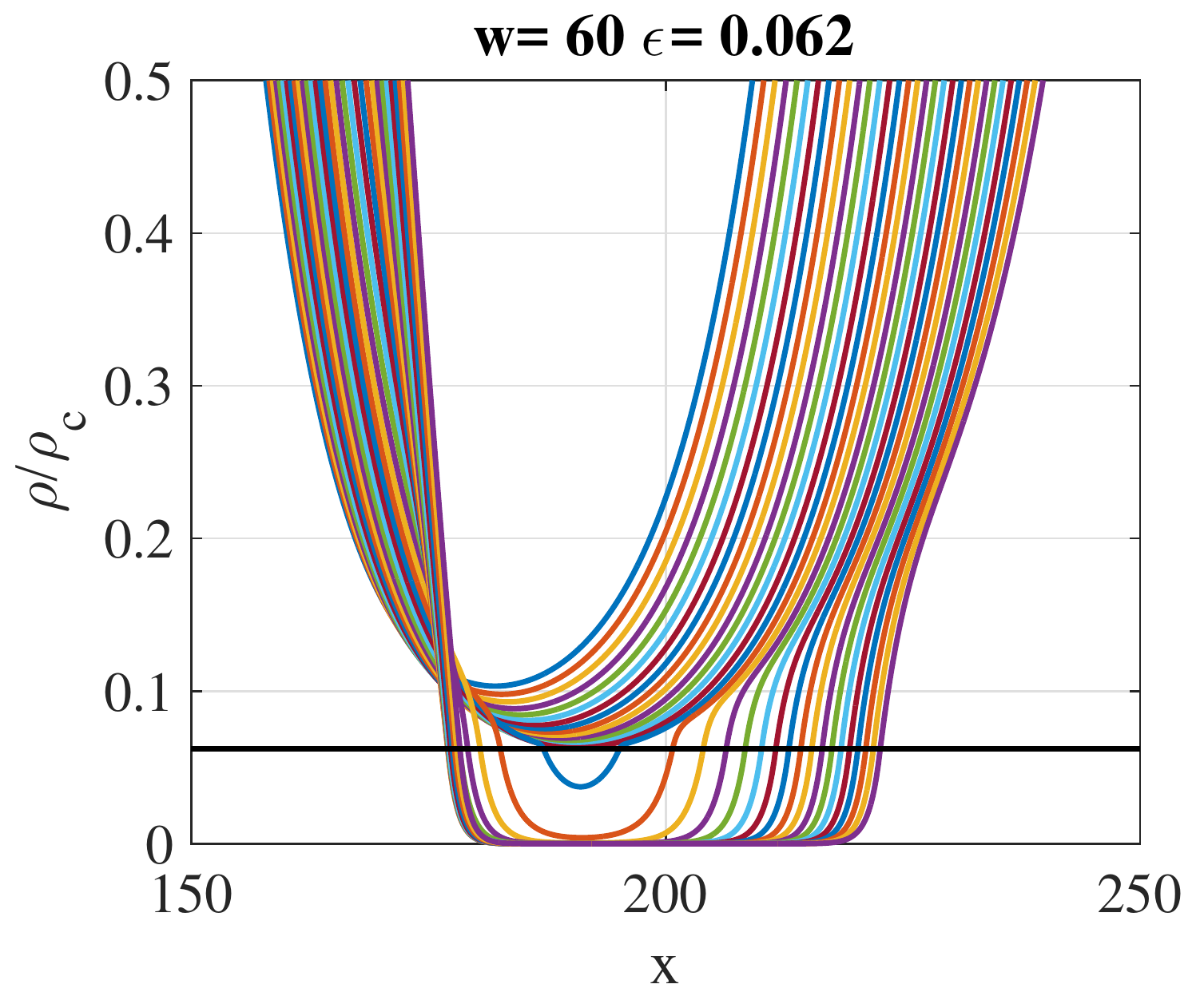}
	\end{subfigure}\hspace*{\fill}
	\begin{subfigure}{0.48\textwidth}
		\includegraphics[width=\linewidth]{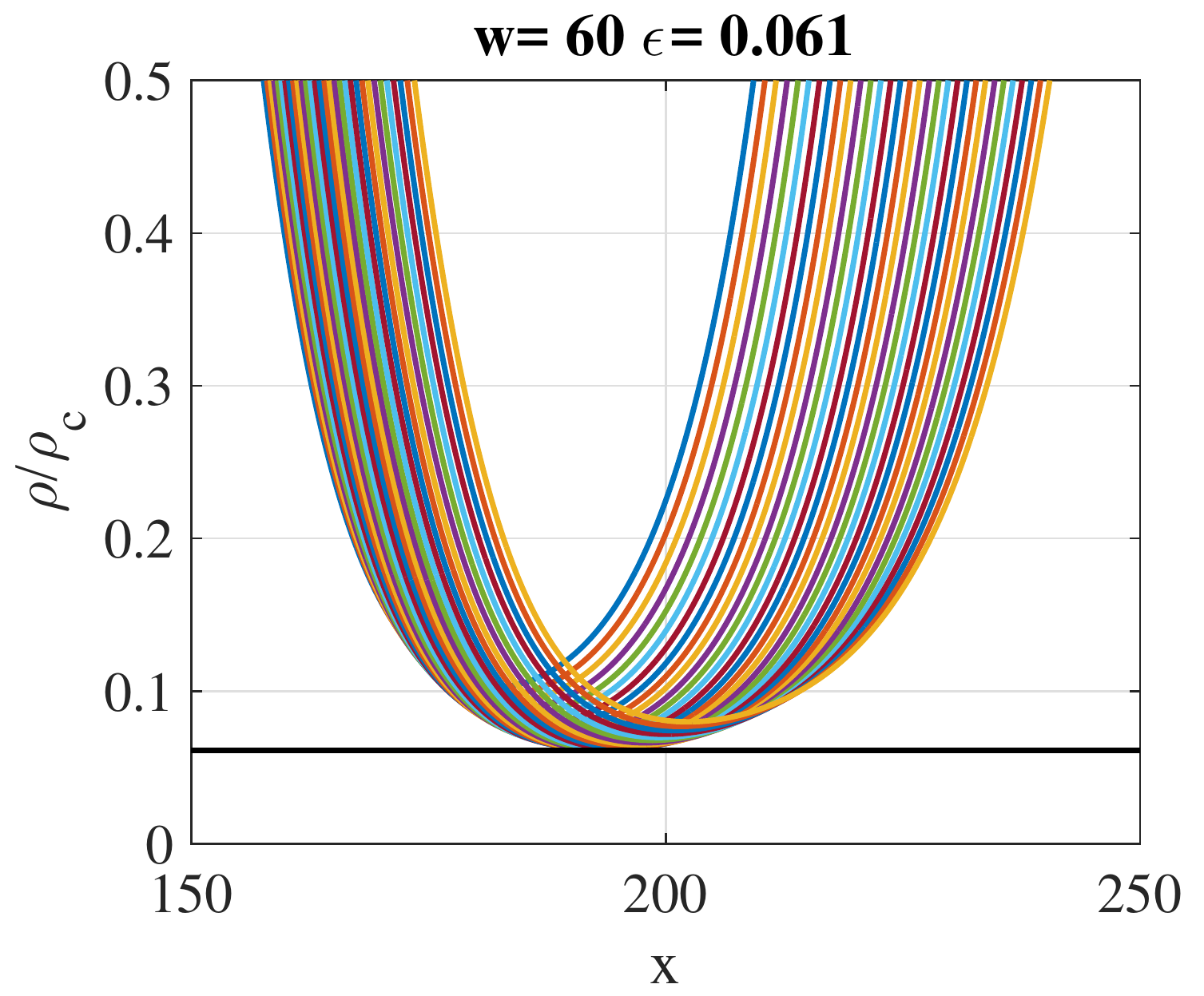}
	\end{subfigure}
	\caption{The population density vs. locations at several times for the Gaussian model. In the left there is separation of species, which does not appear in the right. $D=a=N_c=1$.} 
	\label{fig:SpeciesGdet}
\end{figure*}

In  the case of the Gaussian kernel, the homogeneous solution of the deterministic equation is stable. The appearance of species can only happen because of processes near the propagating front. 
The population density near the front is not monotonic even without cutoff. The non-monotonicity
is a direct result of the nonlocal competition. The population near the front only competes with the population  behind it. Thus the overall competition for it 
is smaller than in the homogeneous state and it results  in a higher population density near the front. The individuals next to that high population density have more competition
than in the homogeneous state, and therefore the population density is lower than the homogeneous state, and so on. Thus, even in the deterministic system,
the propagating state exhibits oscillations in the vicinity of the front, that die out as one gets further into the bulk.
The creation of distinct clusters in the deterministic case with a cutoff for the Gaussian kernel is thus connected to the properties of the first minimum behind the front. If the minimal value of the population density there is
below the cutoff $\epsilon$, the population there declines rapidly.
Figure \ref{fig:SpeciesGdet} presents the area near the first minimum for $w=60$ and two close values of $\epsilon$. Each line represents a different time. 
In the right plot, the population density is always higher than $\epsilon$ and there is no creation of separated species.
In the left plot, the density does dip below  $\epsilon$ and we can see it drops down close to zero immediately afterwards.  The above process, which converts the local minimum near to front to an unpopulated region, is repeated, with the front region recovering its original shape and producing a new local minimum, which is the suppressed, and the cycle continues.  Thus, the separated regions are stationary in space while the front moves with its characteristic velocity.
 If the population density in this minimum is not lower than the cutoff, there is no suppression, and the entire pattern moves forward with the front velocity, with no features stationary in the lab frame.

\begin{figure*}[!ht]
	\centering
	\begin{subfigure}{0.48\textwidth}
		\includegraphics[width=\linewidth]{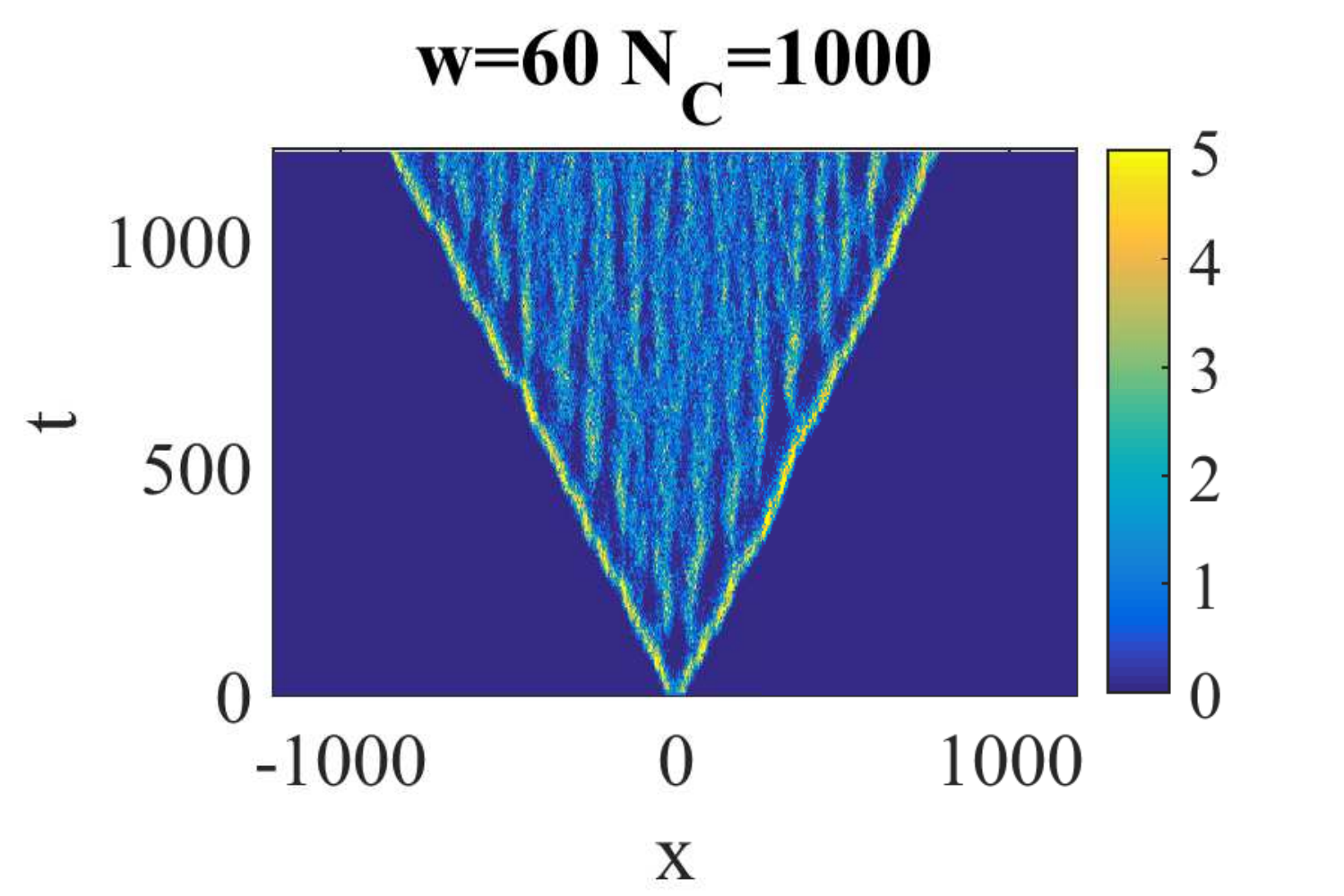}
	\end{subfigure}\hspace*{\fill}
	\begin{subfigure}{0.48\textwidth}
		\includegraphics[width=\linewidth]{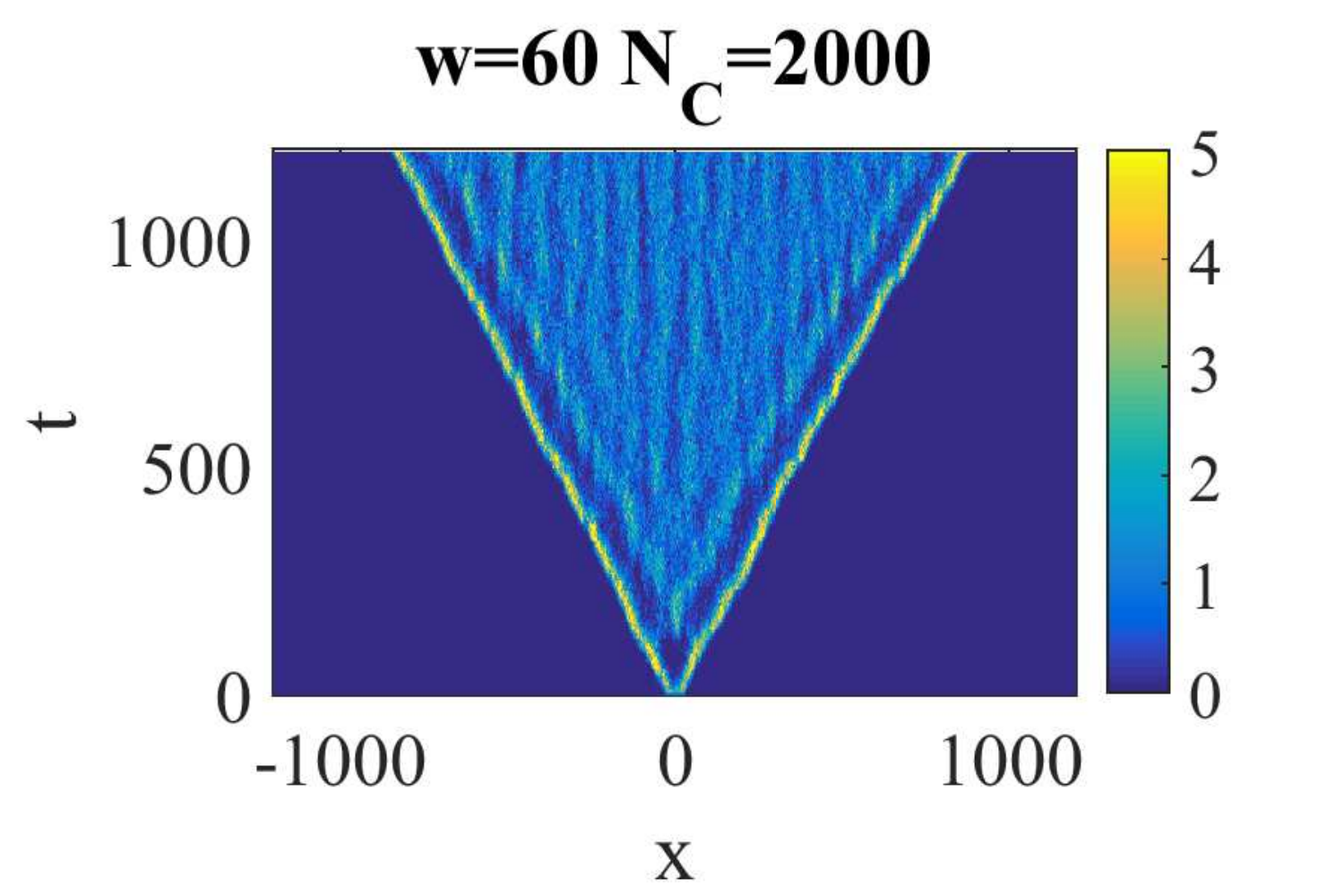}
	\end{subfigure}
	\begin{subfigure}{0.48\textwidth}
		\includegraphics[width=\linewidth]{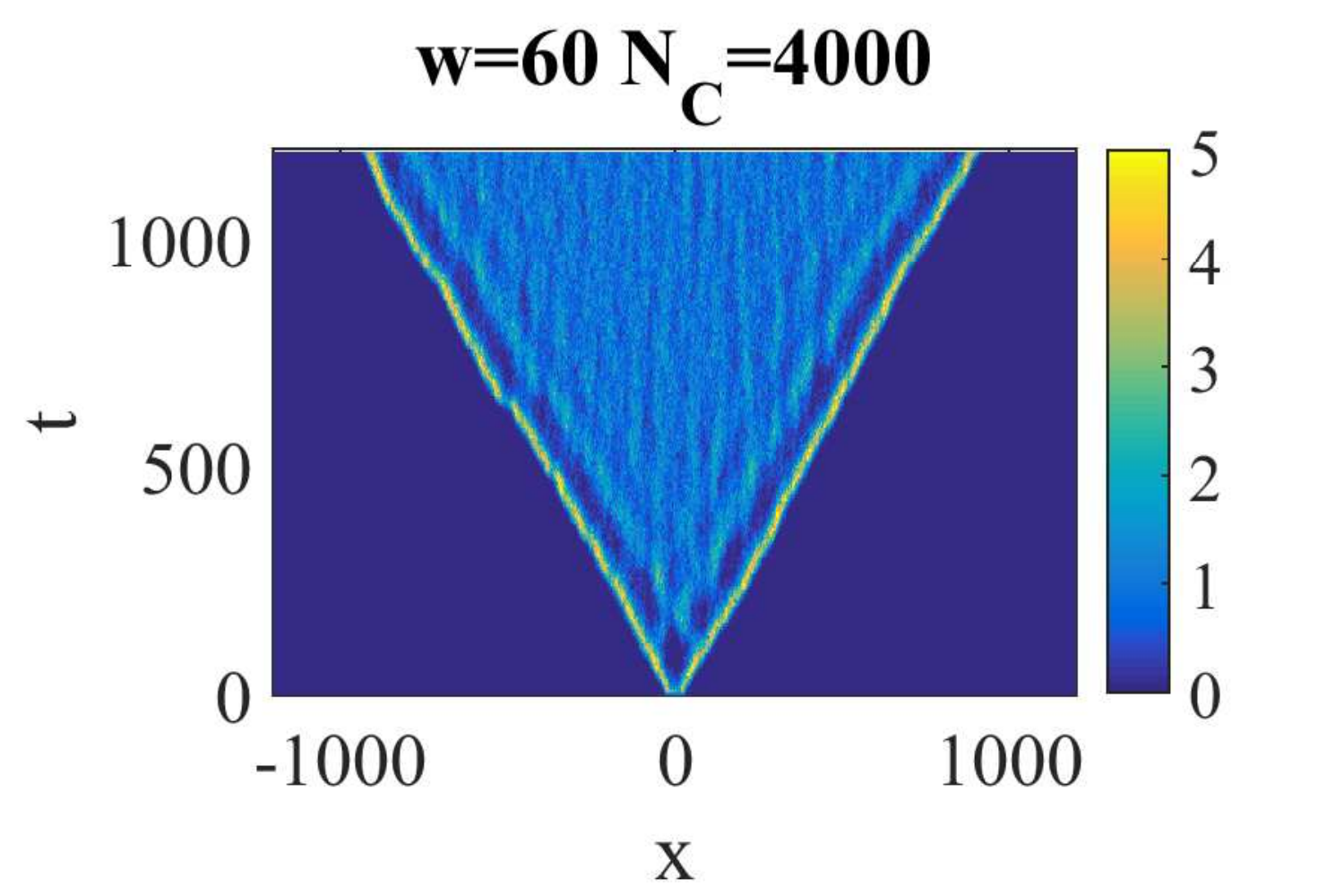}
	\end{subfigure}
	\caption{The time development of the population density for the Gaussian kernel with $w=60$, $N_C=1000$, $2000$, and $4000$. $D=a=1$} 
	\label{fig:Gaussnaps}
\end{figure*}

We have seen that, for this value of $w$,  the critical epsilon is about $\epsilon=0.062$.  Using our phenomenological value of $\pi$ to connect this $\epsilon$ to a value of $N_C$ yields a prediction for the critical value of $N_C$ of $308$ for $w=60$.  For the stochastic simulations, the concept of a critical $N_C$ is problematic.  Nevertheless, there are clear similarities between the deterministic cutoff simulations and the stochastic results.  Most striking is the fact that for large $N_C$ there is a clearly identifiable second peak that propagates together with the leading peak.  For $w=60$, this surely extends down to $N_C$ of 4000, and is even arguably present for $N_C=2000$. For small $N_C$, all but the leading peak are stationary, and have their origin in the leading peak. This is clearly true up to $N_C$ of $1000$ (see Fig.~\ref{fig:Gaussnaps}).  However, the intermediate $N_C$ case has features common to both the ``clustering" and non-clustering phases.  For example, even for $N_C=4000$ there are clusters that are born at the leading edge, and propagate along with the leading peak, and the former second peak interacts with this and becomes stationary.  Thus, it appears that
the most one can say is that the transition in behavior occurs around $N_C=1000-2000$.

\subsection{Flat-top Kernel}
\begin{figure}[!ht]
	\centering
	\includegraphics[width=\textwidth]{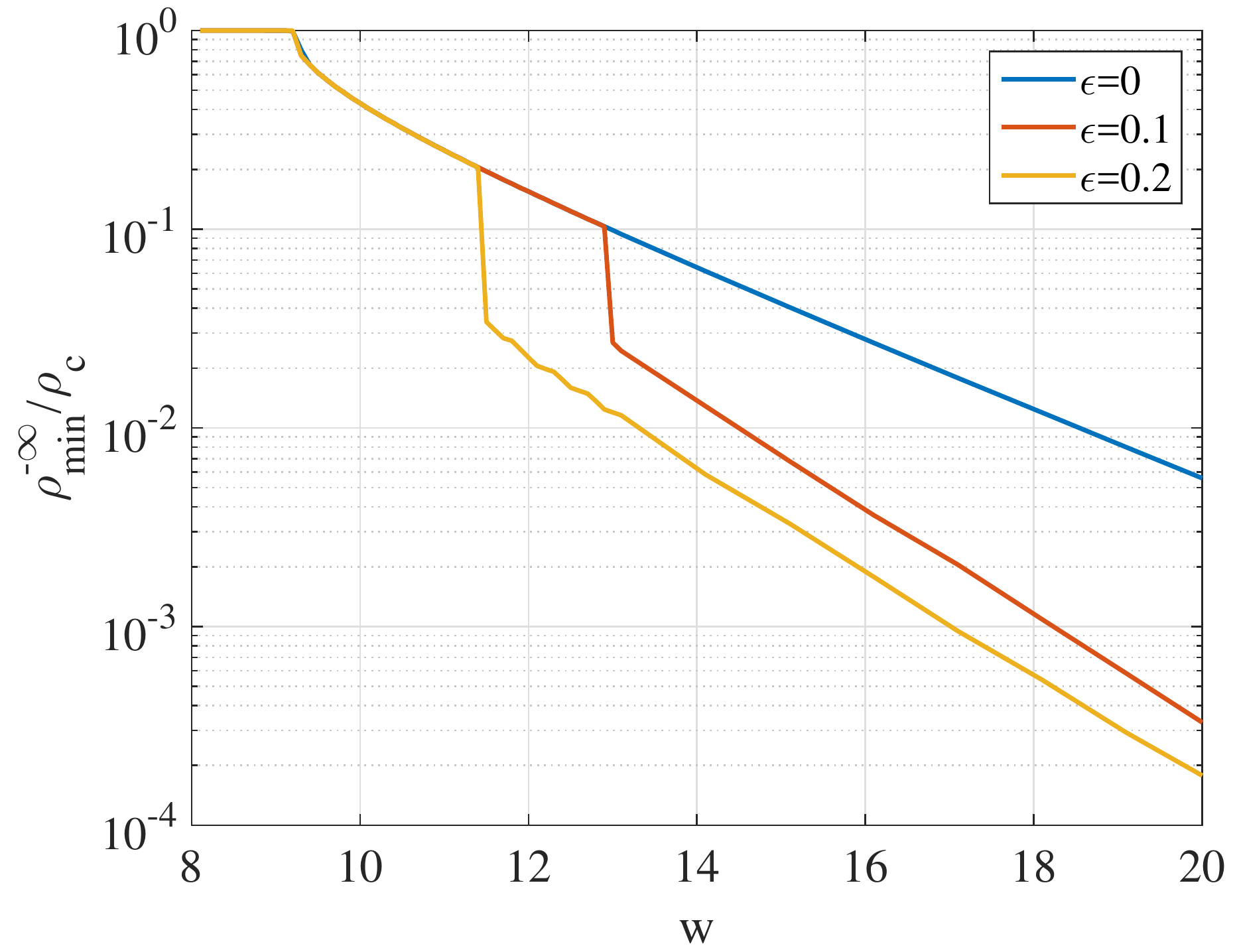}
	\caption{The minimal population density far from the front vs. $w$ for the flat-top model. $D=a=N_c=1$.}
	\label{fig:rhomininf}
\end{figure}

In the case of the flat top kernel, there is a critical value $w_c=9.1761$ (when $a=D=1$), so that for $w>w_c$ the homogeneous deterministic solution is not stable, and patterns emerge even in the absence of stochasticity (or cutoff)
The amplitude of the patterns grows with $w$. When $w$ is relatively big, the oscillations are so large that the population between peaks is very small (but nonzero).  The introduction of the birth cutoff is relevant to the species formation process only when the minimum density between the species 
is lower than the cutoff. This is quite different from the Gaussian kernel where the effect of the cutoff is strongest at the first minimum right behind the front. Figure~\ref{fig:rhomininf} shows the
minimal density far from the front $\rho_{min}^{-\infty}$ vs. $w$ for different cutoffs. We can see that for $w<w_c$ the minimal density is $\rho_c$ since 
the homogeneous pattern is stable. For large $w$ the minimal value drops exponentially. When $\rho_{min}^{-\infty}$ gets below the cutoff we see a rapid drop
and as $w$ keeps decreasing $\rho_{min}^{-\infty}$ decreases exponentially with bigger slope, but two different cutoffs give the same slope. 

\begin{figure}[!ht]
	\centering
        \begin{subfigure}{0.48\textwidth}
                \includegraphics[width=\linewidth]{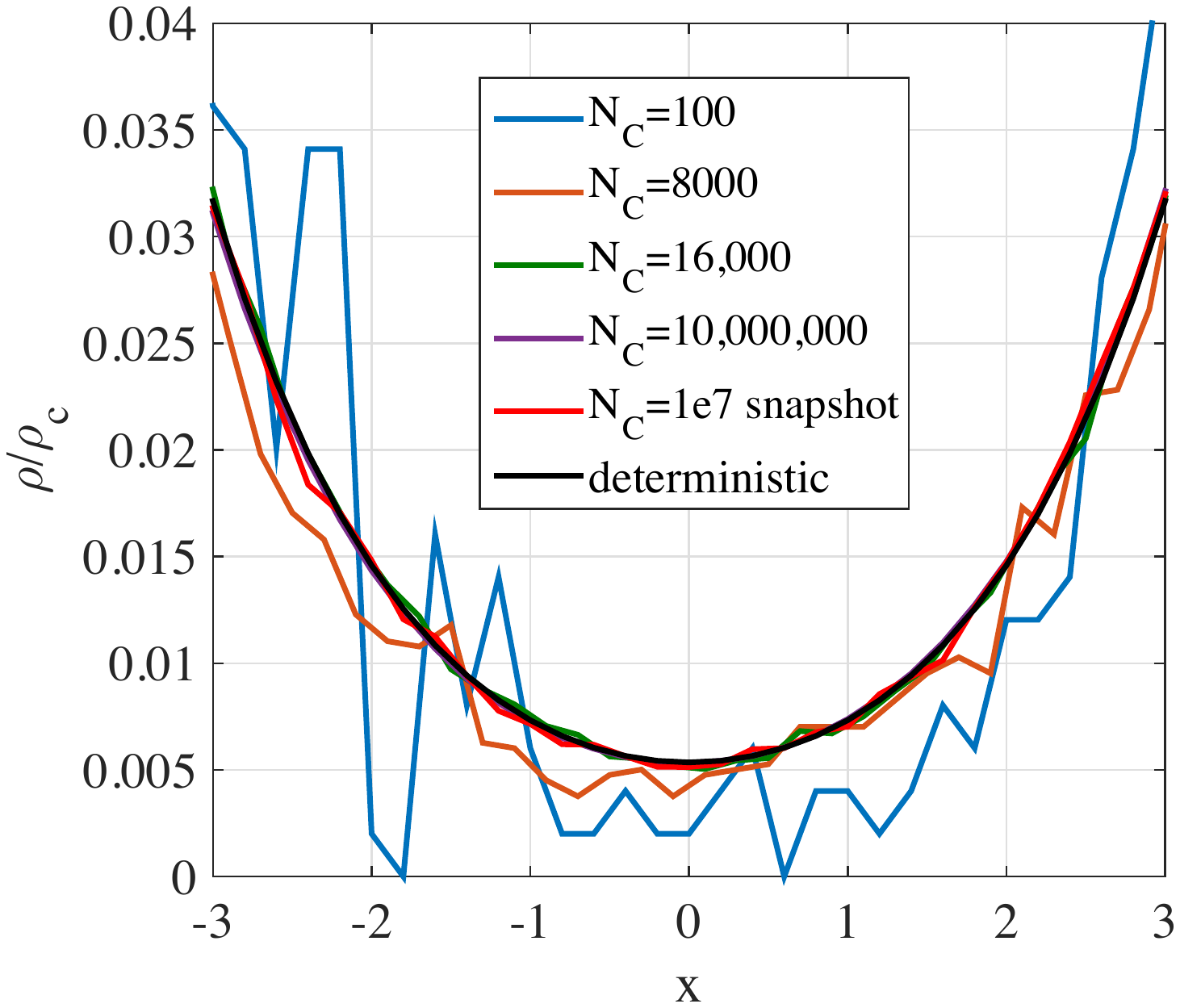}
        \end{subfigure}\hspace*{\fill}
        \begin{subfigure}{0.48\textwidth}
                \includegraphics[width=\linewidth]{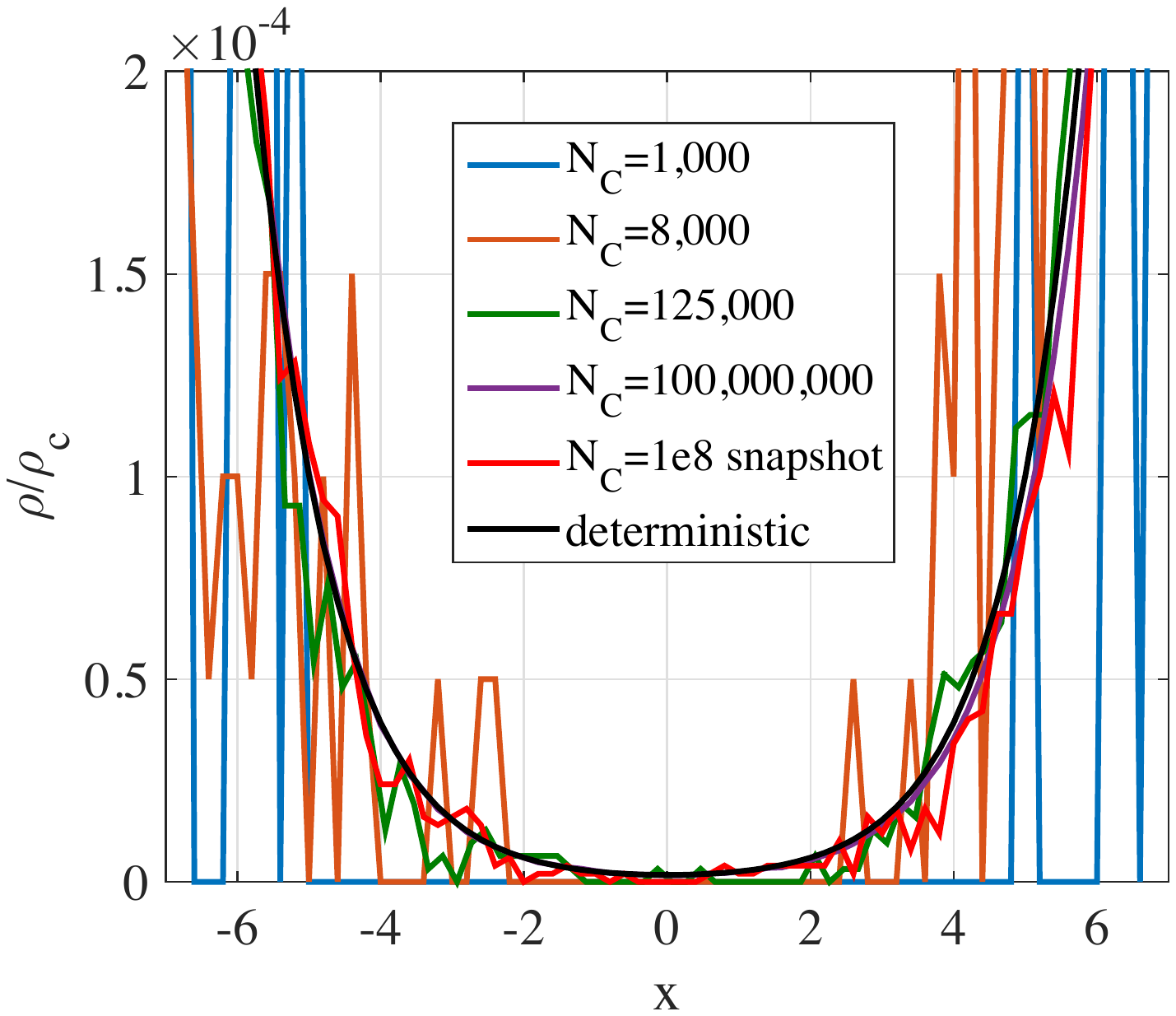}
        \end{subfigure}
        \caption{The normalized population density near the minimal density in the flat top kernel for different values of $N_C$. Left - $w=20$. Right - $w=40$. $D=a=1$.}
        \label{fig:minF}
\end{figure}

The qualitative results obtained in the deterministic simulations of the flat-top kernel hold true for the stochastic case as well. The only difference is that in the stochastic simulation if the population in some 
region dips lower than the birth cutoff it means that there are no individuals there. In fact, there can be some individual that diffuse to there from time to time, but it is a rare event.
Figure~\ref{fig:minF} shows the population density near the minimal density for two values of $w$ and different values of $N_C$. The stochastic simulations are averaged over 500 time snapshots,
and for the largest $N_C$ we also show a non-averaged snapshot. For large $N_C$ there is a good fit between the stochastic simulations and the deterministic case since the birth
cutoff does not applied for these values. For lower values of $N_C$ the population density is lower than the deterministic solution, and even vanishes at times. The discontinuities present for small $N_C$ are due to insufficient samples to accurately measure these small average densities.

\section{Summary}
In this work, we investigated the spatial propagation of finite population in the nonlocal Fisher-KPP equation. In the limit of infinite 
population, the spreading velocity of the population is the Fisher velocity. For large population size, the population spreads through space as 
a wave moving with constant velocity as in the local Fisher equation. As for the local stochastic process, in the nonlocal stochastic process even for fairly large average population, the spreading velocity is much lower than the Fisher velocity. We showed that the main reason behind the slower velocity is the absence of births in the far front, due to the existence of a last leading individual, which can be well captured by introducing a
 cutoff to birth process in the tail of the front.  Adding such a cutoff to the deterministic model brings semiquantitative agreement with the results of 
the stochastic high population simulations. 
In the limit of low population size, the spreading process is different; instead of a front moving with constant velocity, the population now sits in separated clusters
which diffuse in space. The clusters divide at some rate, and each time a cluster divides, the population spreads. The spreading velocity of this process is now 
exponentially slow in the width of the competition, and goes like $v\propto\sqrt{aD}\exp\left(-\frac{b\,\sqrt{a}\,w}{\sqrt{D}N_C^{\alpha}} \right)$. Here $\alpha$
is determined by the competition kernel function. This behavior only happens in the stochastic model; the deterministic model with large birth cutoff shows no spreading
of the population.

We also discussed the appearance of distinct clusters in the large population case - we showed that distinct clusters emerge in the deterministic case upon introducing
the birth cutoff. When the cutoff is larger than the minimal population inside the pattern (as opposed to the tail of the front), the population there declines rapidly.
This brings the separation between species. For the Gaussian kernel, the minimal population density is in the first minimum behind the moving front, this minimum
appears because of the nonlocal competition. For the flat-top kernel, the minimal density is a result of the instability of the homogeneous solution. For this kernel there are
clusters even for infinite population, but introducing a birth cutoff makes the distinction between clusters more clear when the competition is relatively short-ranged. The process of the formation of clusters for high
population in the stochastic model is the same as for the deterministic process.

\bibliographystyle{unsrt}
\bibliography{paper}

\end{document}